# Mapping the Invisible: A Framework for Tracking COVID-19 Spread Among College Students with Google Location Data

Prajindra Sankar Krishnan
Institute of Sustainable Energy
Universiti Tenaga Nasional
43000 Kajang, Malaysia

Chai Phing Chen
Institute of Sustainable Energy
Universiti Tenaga Nasional
43000 Kajang, Malaysia

Gamal Alkawsi
Institute of Sustainable Energy
Universiti Tenaga Nasional
43000 Kajang, Malaysia

Sieh Kiong Tiong
Institute of Sustainable Energy
Universiti Tenaga Nasional
43000 Kajang, Malaysia

Luiz Fernando Capretz
Department of Electrical and Computer Engineering
Western University
London, Canada, N6A5B9

*Abstract*— The COVID-19 pandemic and the implementation of social distancing policies have rapidly changed people's visiting patterns, as reflected in mobility data that tracks mobility traffic using location trackers on cell phones. However, the frequency and duration of concurrent occupancy at specific locations govern the transmission rather than the number of customers visiting. Therefore, understanding how people interact in different locations is crucial to target policies, inform contact tracing, and prevention strategies. This study proposes an efficient way to reduce the spread of the virus among on-campus university students by developing a self-developed Google History Location Extractor and Indicator software based on real-world human mobility data. The platform enables policymakers and researchers to explore the possibility of future developments in the epidemic's spread and simulate the outcomes of human mobility and epidemic state under different epidemic control policies. It offers functions for determining potential contacts, assessing individual infection risks, and evaluating the effectiveness of on-campus policies. The proposed multi-functional platform facilitates the screening process by more accurately targeting potential virus carriers and aids in making informed decisions on epidemic control policies, ultimately contributing to preventing and managing future outbreaks.

*Keywords*— contact networks; human mobility simulation; epidemic control policy

## I. Introduction

Analysis of the spread of infectious diseases, particularly highly contagious ones like COVID-19, heavily relies on understanding human mobility [1-5]. Early identification of potential patients and timely implementation of preventative measures are crucial for slowing the disease's spread and mitigating outbreaks.

To manage epidemics more precisely, understanding individual human mobility becomes essential. The proliferation of devices with accurate localization functions and powerful wireless networks allows for the detection of human mobility over extended periods and large geographical scales. Researchers globally now have access to extensive databases tracking people's movements, enabling detailed analyses and epidemic spread simulations with finer spatial and temporal granularity. GPS and Google location history, classic sources of mobility data, provide valuable information on various types of movements [6].

Traditional epidemic models like Susceptible, Infectious, and Recovered (SIR) [7] or Susceptible, Exposed, Infectious, and Recovered (SEIR) models [8], although foundational, coarsely simulate epidemic tendencies. They assume a constant contact ratio, uniformity across different locations, and simulation for the entire population, neglecting individual nuances. Recent advancements incorporate human mobility into models, overcoming these constraints [9, 10].

This article introduces the Google History Location Extractor, a software platform addressing these limitations with a focus on adaptability and compatibility. The platform leverages actual student trajectory datasets to simulate epidemic spread, estimate on-campus student contact behaviors through GPS trajectories, and evaluate the effectiveness of public policies in slowing down the spread. It features two key functions: a probabilistic model of infectious disease transmission at the individual level and an individual infection risk exploration. Moreover, it enables data mining and exploration of infection spread risks, including the identification of potential secondary contacts.

The proposed platform excels in adaptability to open-world scenarios, accommodating diverse epidemiological situations. It also seamlessly integrates with advanced time series analysis techniques, allowing for a more nuanced understanding of the non-linear behaviors inherent in infectious disease spread. The framework handles various epidemiological scenarios, making it a robust tool for predicting and mitigating COVID-19 spread among university students.

A detailed evaluation of the proposed model's effectiveness underscores its success in real-world scenarios. Through comprehensive simulations and specific metrics, the Google History Location Extractor has demonstrated its capacity to predict and mitigate COVID-19 spread effectively among university students. The platform's adaptability and compatibility contribute significantly to its success.

In terms of mitigation strategies, the Google History Location Extractor offers a valuable tool for policymakers and researchers. It facilitates targeted measures by optimizing contact tracing and assessing the impact of different policies on epidemic dynamics. Policymakers can leverage the platform to make informed decisions, implementing strategies that effectively curb the spread of COVID-19 on university campuses.

This study emphasizes the development of the Google History Location Extractor software as a comprehensive data analysis tool. It provides interactive visualization features for a better understanding of potential COVID-19 spread among university students, contributing to more effective epidemic management and prevention. The remainder of the paper covers modelling mobility patterns, the implementation process, results, and a discussion of our findings. In conclusion, we summarize the key takeaways and emphasize the significance of our platform.

## II. MOBILITY PATTERNS MODELLING SELECTING

This study aimed to analyze mobility data network data of on-campus students, including their latitude and longitude coordinates, visited locations, and proximity between two mobility data devices (within a 1-meter distance) over a 14-day period. In addition, the researchers conducted simulation modeling to investigate mobility patterns and assess the likelihood of COVID-19 transmission.

### A. Direct Contact Model

A direct contact model for COVID-19 was adopted in this study, according to Ghayvat H et al. [12], global mobility network data was utilized to construct a mobility matrix $Mob^{mobility}$, with the matrix's elements represented by $M_{p,q}^{mobility}$. Here, the $M_{p,q}^{mobility}$ denotes the COVID-19 subject's daily mobility movements and the number of individuals who crossed the threshold boundary of 1m and came closer to the COVID-19 subject. It also depicts the COVID-19 subject's relocation movements at a specific location p at a given time t and day d, as well as during the period of incubation (for COVID19, d is set to 15 days) [13]. The direct contact $DC(t, d)$ can be calculated using Eq. (1) [12].

$$DC_{(t,d)} = \frac{\left[\sum_{d=1}^{15}(C19_p^{mobility} x\ \sum_{n=1}^{N}(NS_{n,p}^{mobility})\right] x\ A_{area}}{D_{C19_p, NS_{n,p}}} \quad (1)$$

$C19_{p,q}^{mobility}$ = COVID-19: Mobility at site p;

$NS_n$ = A set of healthy people who travelled close to a COVID-19 subject for a predetermined minute at site p;

n = 1, 2, 3…. N;

d = day, 1, 2, 3…. 15;

t = when the COVID-19 subject and a nearby, healthy person are close to the established threshold limit of 1 m;

n = total number of people who had contact with the COVID-19 subject;

$A_{area}$ = the COVID-19 subject's average mobility and the total number of nearby healthy people depending on the radius $r_{p,q}$ at site p;

$D_{C19_p, NS_{n,p}}$ = distance between a subject with COVID-19 and nearby healthy people.

The parameters mentioned earlier can vary over time, location, and geographical regions. A probabilistic approach is employed by the adopted model [12] to analyze cases of direct contact with COVID-19, utilizing readily available mobility network data. This approach is helpful for tracking contact cases at a specific time and day in a given region. However, the COVID-19 Pandemic Direct Contact Model can be further refined by accounting for individuals who have been in close proximity to a COVID-19 patient for a predefined duration of minutes or more. To achieve this, an Indirect Contact Model [11] was also incorporated, which enables the identification of individuals who may have been indirectly exposed to COVID-19.

### B. Indirect Contact Model

To analyze the dynamics of indirect COVID19 transmission, a probabilistic COVID19 pandemic indirect contact model is employed to identify individuals who have indirectly come into contact with COVID-19. To achieve this, Ghayvat H, et al. [12] constructed an indirect mobility matrix, $iM^{mobility}$, using global mobility network data, with $M_{p,q}^{mobility}$ as its element. $M_{p,q}^{mobility}$ shows the everyday moves of healthy people who have been in proximity to COVID-19 exposed individuals for more than the predefined duration of minutes in a particular area [14]. In addition, it considers their relocation moves between the areas p and q, as recorded for a period of 15 days, starting from time t, day d, and last day d-1. The indirect infection suspicion $ID_{(t,d)}$ can be calculated using Eq. (2) [12].

$$ID_{(t,d)} = \frac{\left[\sum_{d=1}^{15}(iC19_p^{mobility} x\ \sum_{n=1}^{N}(iNS_{n,p}^{mobility})\right] x\ A_{area}}{D_{iC19_p, NS_{n,p}}} \quad (2)$$

$iC19_{p,q}^{mobility}$ = suspected COVID19 subject Mobility at location $p$;

$iNS_n$ = a list of nearby healthy people who have travelled near a probable COVID-19 subject for a predefined duration of minutes at location $p$;

$D_{iC19_p, iNS_{n,p}}$ = distance between suspected COVID-19 subjects and the neighbouring healthy individuals.

The distance between two individuals ($D_{C19_p, NS_{n,p}}$ and $D_{iC19_p, iNS_{n,p}}$) at a specific time is determined by applying the Haversine formula, which utilizes the latitude and longitude information of the two individuals' locations to determine the distance between them in kilometers. The formula can be expressed as follows:

$$a = sin^2\left(\frac{\Delta lat}{2}\right) + cos(lat1) \times cos(lat2) \times sin^2(\frac{\Delta long}{2}) \quad (3)$$

$$c = 2 \times arctan2(\sqrt{a}, \sqrt{1-a}) \quad (4)$$

$$d = R \cdot c \quad (5)$$

Distance in km based on Pythagoras' theorem is given by equation:

$$x = (long2 - long1) \times cos(\frac{lat1+lat2}{2}) \quad (6)$$

$$y = lat2 - lat1 \quad (7)$$

$$d = R(\sqrt{x \times x + y \times y}) \quad (8)$$

Where

$R$ = Earth's radius (6371 km)

$lat1, long1$ = latitude, longitude pair of 1st point

$lat2, long2$ = latitude, longitude pair of 2nd point

$\Delta lat$ = difference between two latitudes

$\Delta long$ = difference between two longitudes

$c$ = Axis interaction calculation

$d$ = distance in kilometres

### III. IMPLEMENTATIONS

#### A. Mobility Data

The ubiquity of smartphones has provided a tremendous opportunity to collect large amounts of location data from individual users on an unprecedented scale. Human mobility is a crucial element in understanding various disaster events. In the case of infectious diseases, the transmission occurs between humans, and knowledge of human mobility patterns can help epidemiologists predict disease outbreaks. Therefore, understanding how people move through time and space, which can be done with the help of mobile phone location data, is crucial to responding to and recovering from such disaster events. One of the key components that describe these mobilities is location history. An individual's geographical location can be determined via different methods [15], such as:

*a) Cell site location method:* using triangulation techniques, this method determines the location of a mobile phone by figuring out how far apart two or three different signal antennas are from one another. Every mobile phone has a connection to a cell tower for communication, and whenever a phone travels, it sends a ping to the closest signal tower to provide its current position.

*b) Global Positioning System (GPS):* GPS technology offers 3D locations quickly and constantly when satellites are available. The GPS in mobile phones enables position tracking through a satellite system. From time to time, these places assist in tracking user activities. Google Maps primarily uses GPS to show daily life activities in the user's google timeline history by accessing the phone's GPS and continuously tracking all activities.

*c) Wi-Fi positioning:* If the mobile device's Wi-Fi is enabled, it always looks for Wi-Fi connections. The individual can be found by using the location of close Wi-Fi. However, this technique is not very effective for locating a person.

Google Maps employs all three of these technologies —cell site location, GPS, and Wi-Fi positioning—to accurately locate the user's location.

The timeline feature of Google Maps provides a concise overview of a user's location history. To export this data, users can utilize the Google Takeout service, which provides the data in JSON format. The exported JSON file contains a vast number of metadata, with certain data being more significant than others. Outlined below are some of the key data points.

- Heading: Provides information regarding the device's current travel direction.
- Activity type: Describes the kind of action performed by the user. For example, walking, still, etc.
- LatitudeE7: Provides latitude information in 9 digits.
- LatitudeE7: provides longitude information in 9 digits.
- Accuracy: Shows the degree of accuracy of the data item. Typically, accuracy measured in units less than 800 is regarded as high precision, whereas accuracy measured in units more than 5000 is regarded as poor accuracy.
- Timestamp MS: a 13-digit code that provides millisecond-level information on the date and time of an event's occurrence.
- Altitude: It provides data about elevation above sea level.

Google Maps provides a range of functionalities in addition to searching for places and finding routes. Google Timeline, a component of Google Maps, constantly tracks user activities, including the places visited and the mode of transportation used to get there, through GPS location data. This raw data describes the user's location at a specific date and time.

In conducting this study, ethical considerations were prioritized to ensure the responsible handling of sensitive location history data. The research involved a cohort of 50 volunteer university students who willingly shared their Android smart mobile location history for the purpose of advancing knowledge in the field. These students were informed about the nature and scope of the study, emphasizing the voluntary nature of their participation and the confidentiality of their data.

The data collection period spanned 14 days, from April 14 to April 28, 2022. This timeframe was chosen to capture a representative snapshot of the participants' mobility patterns while minimizing potential bias from seasonal variations. Notably, only students residing on the university campus were included in the pilot study. This intentional focus aimed to investigate the potential spread of COVID-19 among university students in close living quarters, contributing to a targeted and contextually relevant examination. Figure 1 illustrates the Venn diagram used to analyze the location data of two students.

Demographic information, such as age, gender, and academic major, was collected alongside the location data to enhance the transparency of the research methodology. This additional context allows for a more comprehensive understanding of the participants and potential variations in behavior or mobility patterns within the studied population.

The location histories of the selected students underwent rigorous analysis utilizing the Google History Location Extractor and Indicator. The methodology involved plotting the students' locations on a map, providing a visual representation enriched with comprehensive details, including date, time, region, and the calculated distance between individuals. This meticulous analysis aimed to identify potential scenarios of COVID-19 spread, contributing valuable insights to the study's objectives.

Throughout the research process, ethical guidelines, and data protection principles were strictly adhered to, ensuring the privacy and confidentiality of the participants. The study's findings are presented with due consideration to these ethical standards, promoting transparency and trust in the research methodology.

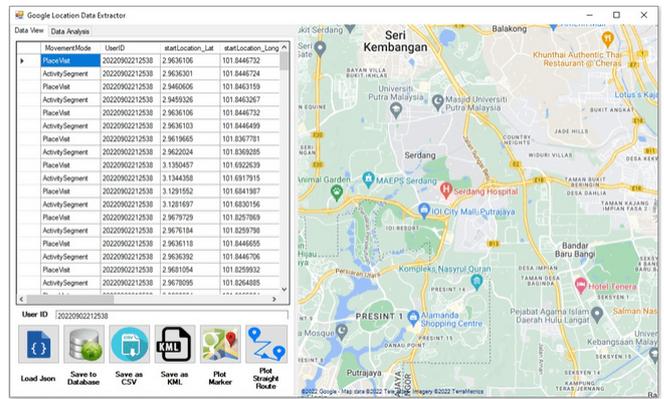

Fig. 2. GUI of Google Historical Location Extractor software.

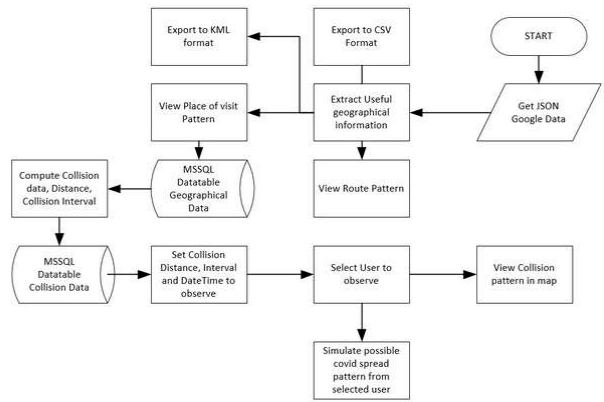

Fig. 3. Entire system architecture.

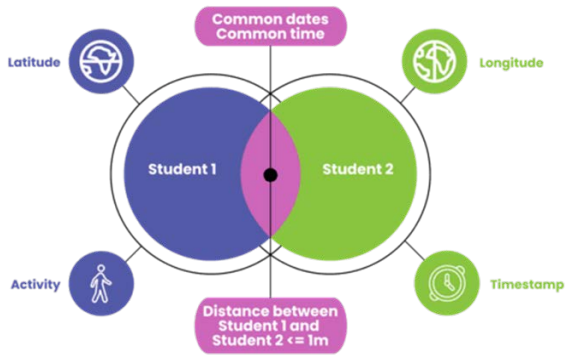

Fig. 1. Venn diagram of location analysis for two-students.

### B. Software Implementation

This study delves into the analysis of location history data from 50 university students, leveraging the capabilities of the "Google History Location Extractor" framework. This standalone tool, showcased in Figure 2, stands out for its versatility in data extraction, graphical user interface display, and mobility visualization on Google Maps. Furthermore, it supports data export in various formats, including .csv or .kml. It allows for advanced analysis by exporting to Microsoft SQL Server database format, as depicted by the entire system architecture in Figure 3. Figure 4 depicts the overall algorithm flow visualization. The following integration of additional points accentuates the framework's adaptability and compatibility.

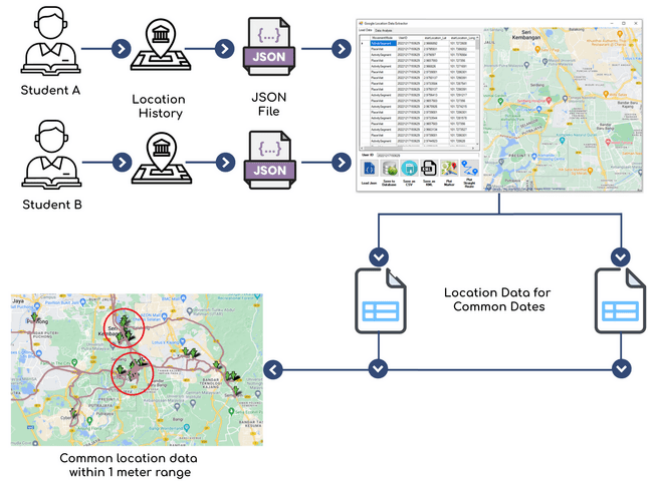

Fig. 4. General flow of algorithm

*a) Data Collection:* The framework exhibits robust capabilities in handling open-world scenarios by collecting data from 50 distinct Android smartphones. These devices, voluntarily contributed by university students over a 2-week period, as shown in Figure 5, offer a dynamic and heterogeneous representation of human mobility patterns. This diverse dataset underscores the adaptability of the framework to various epidemiological cases, providing a foundation for comprehensive analysis beyond the immediate university campus.

Fig. 5. Extracted Data Display

*b) Data Importation and Filtering:* The adaptability of the framework is further evident in its data importation and filtering phases. Utilizing Google Timeline for data collection in JSON format, the framework ensures seamless integration with different data sources, facilitating the analysis of diverse scenarios. The importation process, implemented in VB.net, transforms the JSON file into a structured table format, accommodating various data types. This flexibility not only supports the incorporation of sophisticated time series analysis techniques but also ensures the framework's adaptability to diverse epidemiological situations.

*c) Advanced Time Series Analysis:* The compatibility of the framework with advanced time series analysis techniques is highlighted in the data importation phase. The software's flexibility in handling various data types, including text, dates, and integers, paves the way for the incorporation of sophisticated time series analysis. This adaptability is crucial for capturing non-linear behaviors inherent in epidemiological historical series, making the framework a robust tool for nuanced analysis.

*d) Distance Calculation and Visualization:* The framework's adaptability is further exemplified in the distance calculation stage, where the 'haversine' formula is employed to determine proximity based on latitude and longitude as shown in Figure 6. This method enhances compatibility with data-driven time series analysis, providing a robust approach to assess interactions between individuals. The visualization of results on a map, facilitated by Google's Direction API, not only communicates potential COVID-19 spread scenarios effectively but also signifies the framework's adaptability in presenting complex analysis outcomes.

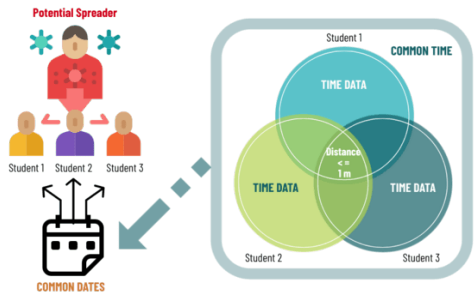

Fig. 6. Proximity analysis for multiple contacts.

The "Google History Location Extractor" framework excels in addressing open-world scenarios by accommodating diverse epidemiological cases. Its compatibility with advanced time series analysis techniques ensures a nuanced understanding of non-linear behaviors in infectious disease spread, making it a valuable tool for comprehensive epidemiological studies beyond the specific context of university campuses.

## IV. RESULTS

In this section, the individual student detection method will be elaborated based on the SEIR model, which includes a deterministic compartment model for the spread of an infectious disease, which serves as the basis for the investigation of infection risk and is further explained in the subsections that follow.

### A. SEIR model

The SEIR model [16], divides a population into four distinct groups based on their stage in the disease progression. The first group is the susceptible population, which is at risk of acquiring the disease. Since SARS-CoV-2 is a new virus, it is assumed that everyone who has not been infected before is susceptible. The second group is the exposed population, which has been exposed to the virus but is not yet contagious. The third group is the infectious population, which has contracted the virus and can spread it to others. The final group is the recovered population, which has recovered from the infection and is no longer susceptible to the disease.

The compartment model can be diagrammed as follows:

$$Susceptible \xrightarrow{\frac{\beta SI}{N}} Exposed \xrightarrow{\alpha E} Infectious \xrightarrow{\gamma I} Recovered$$

The rate processes are modelled as follows [16]:

The rate at which the susceptible population comes into contact with the infected population and the illness is transmitted is known as the $\frac{\beta SI}{N}$. The susceptible population's size is $S$. The model's parameters $\beta$, expressed in units of 1 day.

$\alpha E$ is the rate at which an exposed population becomes infective, where $E$ is the exposed population size. The average time spent in the exposed condition is the disease's incubation period, which is equal to $\frac{1}{\alpha}$.

$\gamma I$ represents the rate at which the infected population recovers and develops resistance to future infection. $I$ represents the infectious population size.

The deterministic SEIR equations provide an essential model for the propagation of an infectious illness in a uniform population. This leads in a system of four equations after substitution [16]:

$$\frac{ds}{dt} = -\beta si$$

$$\frac{de}{dt} = \beta si - \alpha e$$

$$\frac{di}{dt} = \alpha e - \gamma i$$

$$\frac{dr}{dt} = \gamma i$$

where $s + e + i + r = 1$ is an invariant.

Figure 7 is presented without the exposed population compartment, whereas Figure 8 shows the inclusion of an exposed population compartment, which slows the spread but does not seem to lower the number of students infected with the disease. The student population size is set to 50.

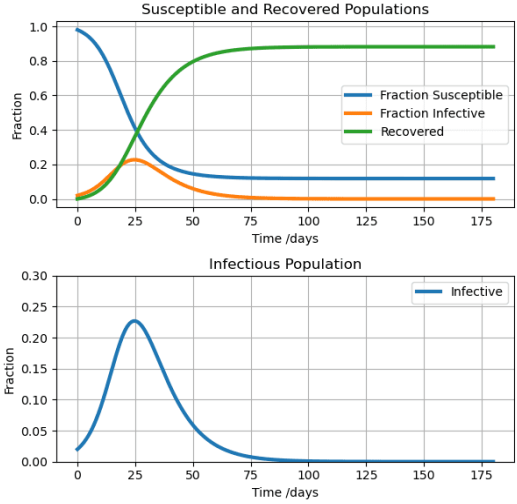

Fig. 7. Simulation of the SIR model

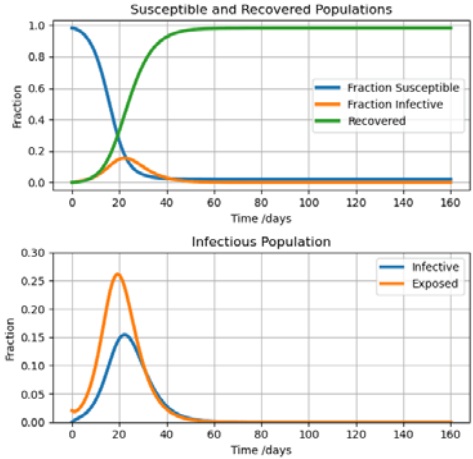

Fig. 8. Simulation of the SEIR model

The goal of campus policies is to decrease the transmission of the virus through social distancing measures that aim to prevent infective individuals from spreading the virus to susceptible individuals. To incorporate the effectiveness of these measures in modeling, a control parameter called μ is introduced, with a value of 0 indicating no control and a value of 1 indicating perfect isolation of infective individuals. The model aims to investigate how the implementation of social distancing measures can influence the outcome of an epidemic.

Exposed. The subpopulation that has been exposed to the disease but is not yet infective.

The compartment model can be diagrammed as follows [16].

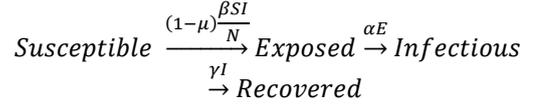

$$Susceptible \xrightarrow[\gamma I]{(1-\mu)\frac{\beta SI}{N}} Exposed \xrightarrow{\alpha E} Infectious \to Recovered$$

The rate processes are modeled as follows:

The rate at which a susceptible population contracts a disease from the infected population is represented by $(1 - \mu)\frac{\beta SI}{N}$. The effectiveness of public health interventions in controlling disease transmission is described by μ, where μ = 0 signifies the absence of effective interventions, and μ = 1 indicates complete elimination of disease transmission.

After substitution, this results in a system of four equations [16]:

$$\frac{ds}{dt} = -(1-\mu)\beta si$$

$$\frac{de}{dt} = (1-\mu)\beta si - \alpha e$$

$$\frac{di}{dt} = \alpha e - \gamma i$$

$$\frac{dr}{dt} = \gamma i$$

where $s + e + i + r = 1$ is an invariant.

The impact of on-campus mobility restriction policies in slowing down the spread of the epidemic, reducing the fraction of infected students, easing the burden on health care resources, and decreasing the number of students eventually contracting the disease, as well as saving lives for diseases with non-zero mortality, is depicted in Figure 9.

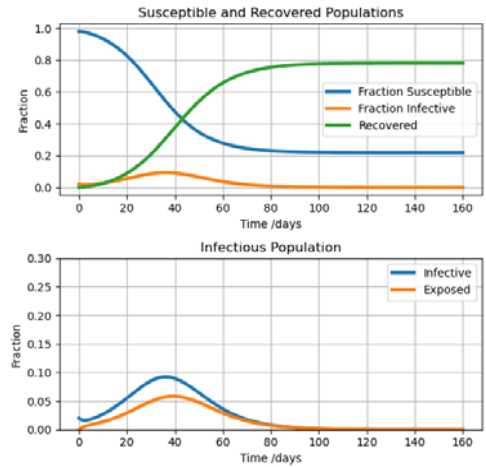

Fig. 9. Simulation of the SEIR model with Control Parameter, μ

### B. Infection Risk Exploration

This study aimed to develop a GUI system for detecting potential contacts between students on campus based on their travel history using data analysis. To achieve this, the process of capturing and verifying individual infections is facilitated through a multi-step approach embedded in the developed GUI system. The system incorporates a tracing function aimed at categorizing contacts into distinct risk levels based on the chronological order of their interactions. This categorization proves crucial for stratifying screening and quarantine strategies, particularly in situations where resources are constrained. The visualization of students' movements, as illustrated in Figure 10 using Google Mobility Data and Google Direction API, provides a tangible representation of the routes and locations frequented by different students on campus.

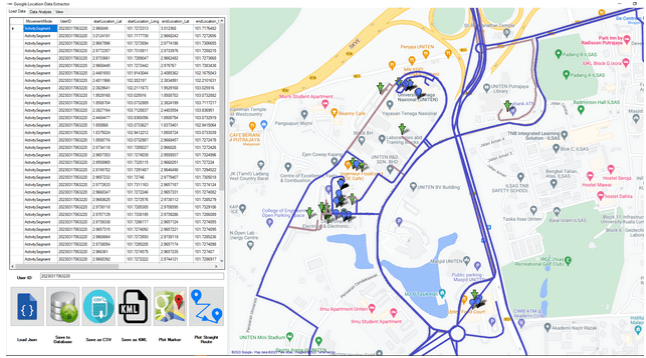

Fig. 10. Location history of students displayed on the same map.

Following this, Figure 11 presents the results of common location data analysis, focusing on eliminating uncommon locations and marking the campus location for a more streamlined data processing approach. The subsequent step involves configuring proximity analysis settings in the Data Analysis tab, as showcased in Figure 12. Users can input various parameters such as Start Date, Start Time, Interval, Collision Distance, Collision Interval, and select a specific user for proximity analysis. These settings serve as criteria for identifying potential contacts and risk stratification.

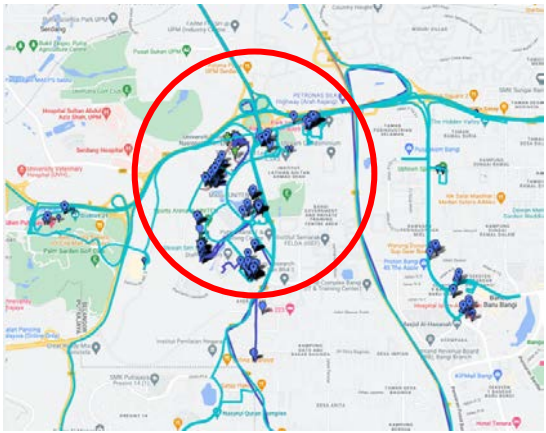

Fig. 11. Common student locations on campus

The result table in Figure 12 then details information such as date, time, latitude, longitude, visited location, and levels of contact. Contacts are categorized into different levels, with Level 1 denoting direct contact, Level 2 representing indirect contact from Level 1, and Level 3 indicating indirect contact from Level 2. The infected user, selected under the Collision User option, is surrounded by multiple susceptible individuals across these contact levels.

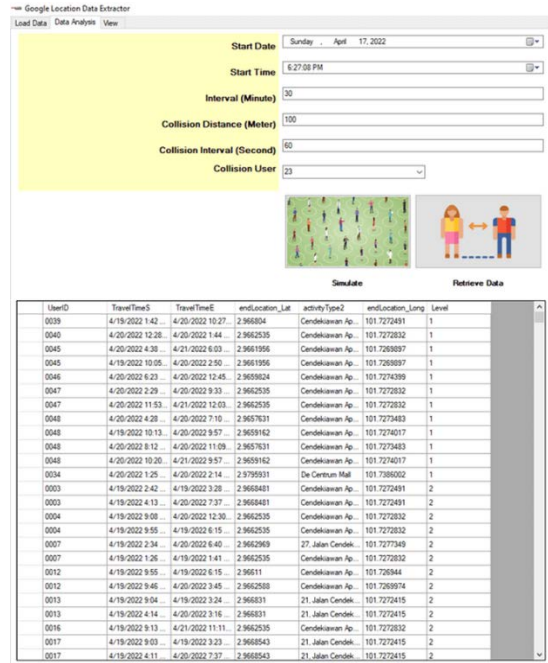

Fig. 12. GUI settings for proximity analysis

This approach enables the prioritization of screening efforts, acknowledging the exponential growth in the number of contacts with each level. Primary contacts (Level 1), having the highest risk, are screened first, followed by secondary (Level 2) and tertiary contacts (Level 3). This sequential screening strategy optimizes resource allocation, ensuring a targeted and efficient approach to identifying and verifying potential infections. In essence, the GUI system provides a robust framework for capturing and verifying individual infections by integrating visual movement analysis, risk categorization, and a prioritized screening strategy based on varying levels of contact.

## V. Discussion

Insights into COVID-19 transmission dynamics are crucial for developing effective mitigation strategies. This study aimed to enhance our understanding of the complex interactions among university students and their potential impact on virus spread. The mobility networks analysis, utilizing the proposed "Google History Location Extractor," provided valuable information about potential transmission locations and links between nodes. The framework's adaptability to open-world scenarios was evident in its analysis of location history data from 50 distinct Android smartphones, capturing a dynamic representation of human mobility patterns in a diverse university setting.

The compatibility of the framework with advanced time series analysis techniques was seamlessly integrated during the data importation phase. Implemented in VB.net, the software transformed the JSON file into a structured table format, accommodating various data types. This flexibility enables the incorporation of sophisticated time series analysis techniques, surpassing the limitations of traditional modelling such as SIR and SEIR, and capturing the non-linear behaviours inherent in epidemiological historical series.

Acknowledging the limitations, incomplete mobility data resulted in exclusions from the network analysis. Despite this, the developed visualization tool proved instrumental in identifying algorithm-detected cross-paths and recent contacts. The absence of data from uninfected individuals or adherence to mitigation strategies is recognized as a limitation, emphasizing the need for comprehensive data collection strategies.

## VI. CONCLUSIONS

This study introduces a data analysis method employing a graphical user interface (GUI) to identify common travel history and predict violations of social distancing among students. The GUI, a component of the "Google History Location Extractor," facilitates mapping analyzed information on a geographical interface, providing essential details for contact tracing. Region mining integrated into the GUI enables fast estimation of potential risks and identification of infections at an early stage, serving as a valuable tool for organizations like the center for disease control and prevention.

The GUI's advantages, such as predicting visits to high-risk areas and analyzing large amounts of student data simultaneously, position it as a valuable asset in tracking and preventing COVID-19 spread. However, considerations for battery life and constant internet connectivity are noted. Further research is warranted to assess the method's feasibility and effectiveness as part of a comprehensive mitigation strategy, with a focus on its adaptability to diverse epidemiological scenarios and compatibility with advanced time series analysis techniques.

## ACKNOWLEDGEMENT

This work is supported by IDRC Grant No. 109586 – 001 for Artificial Intelligence Framework for Threat Assessment and Containment for COVID-19 and Future Epidemics while Mitigating the Socioeconomic Impact to Women, Children and Underprivileged Groups.